\documentclass[aps,prl,groupedaddress,amsmath,amssymb,amsfonts,showpacs,twocolumn,floatfix,a4]{revtex4}

\usepackage{bm}
\usepackage{graphicx}
\usepackage{color}

\usepackage{epsfig}

\begin{document}

\title{Superharmonic Josephson relation at $0$-/$\pi$-junction transition}

\author{M. Houzet$^1$, V. Vinokur$^2$, and F. Pistolesi$^{3,2}$}

\affiliation{
$^1$Commissariat \`{a} l'\'{E}nergie Atomique, DSM/DRFMC/SPSMS, 38054 Grenoble, France \\
$^2$Argonne National Laboratory, 9700 S.Cass Ave, Argonne, IL 60439, USA\\
$^3$Laboratoire de Physique et Mod\'elisation des Milieux
Condens\'es, CNRS-UJF B.P. 166, F-38042 Grenoble, France
}

\date{\today}


\begin{abstract}
Critical current was recently
measured near the transition from $0$
to $\pi$-contact in
superconductor/ferromagnet/superconductor Josephson junctions.
Contrary to expectations, it does not vanish at the
transition point.
It shows instead a tiny, though
finite, minimum.
The observation of fractional Shapiro steps reenforces the idea that
the vanishing of the main sinusoidal term in the Josephson relation
gives room to the next harmonics.
Within quasiclassical approach we calculate the Josephson relation
taking into account magnetic scattering.
We find that the observed minimum is compatible with the value of
the second harmonics expected from the theory.
\end{abstract}

\maketitle

%\section{Introduction}

According to textbooks equilibrium supercurrent, $I$,  in a
tunnel-barrier Josephson junction depends sinusoidally on the
phase difference, $\phi$, between the superconducting leads:
$I=I_c \sin \phi$, where $I_c>0$ is the so-called critical current
of the junction.  In his seminal work of
late seventies Bulaevskii {\it et al} predicted~\cite{bulaevski}
that the sign of $I_c$ can change (or, equivalently, a shift of
$\pi$ appear in the argument of the sine) in the presence of
magnetic impurities within the tunnel barrier.
Soon after, Buzdin {\em et al.}~\cite{buzdin,buzdin2} suggested
that such a junction, conventionally called now the
$\pi$-junction, can be realized in a hybrid structure where the
tunnel barrier is replaced with the ferromagnetic metal.
While predicted theoretically, experimental realizations of
$\pi$-junctions remained long unobserved.
Indeed, superconductivity and magnetism compete; thus conventional
ferromagnets would strongly suppress supercurrent.
The first recent successful realization of a $\pi$-junction
~\cite{ryazanov,kontos} utilized the so-called {\em weak}
ferromagnets,
and the observation of a non-monotonic dependence of $I_c$ as a
function of the temperature \cite{ryazanov,sellier} and on the
thickness of the ferromagnetic layer \cite{kontos} served as the
first evidences of the actual realization of $\pi$-junctions.
Moreover in the former case, the existence of the temperature
$T^*$, where $I_c$ reached a minimum with the vanishing magnitude,
allowed for a precise identification of the transition point.

These spectacular observations of $\pi$-junction behaviors
remarkably confirmed original predictions
of~\cite{bulaevski,buzdin,buzdin2} and yet posed new puzzling
questions.
The first was that the observed amplitude of the current in the
junctions appeared two orders of magnitude smaller than that
expected from the theory.
H. Sellier {\it et al}~\cite{sellier} proposed that magnetic
impurities in the ferromagnet could be the origin of this effect,
and indeed it was recently shown for the $\sin\phi$ 
component~\cite{ryazanovbuzdin} that magnetic
impurities can lead to a noticeable reduction of the critical
current if one assumes somewhat artificial uniaxial distribution
of magnetic disorder.
It remains however to understand the effect of a more realistic
disorder distribution.

Another puzzle concerns the form of the phase-current relation at
the transition point $T^*$.
Quite generally, the phase relation has to be periodic in $\phi$.
This does not rule out the possibility of the second or even
higher harmonics: $I=I_1\sin\phi+I_2\sin2\phi+\dots$, which
indeed appear in Josephson junctions formed by point contacts,
constrictions, or non-equilibrium normal metals~\cite{golubov,note}.
Only in the non-equilibrium case was the $I_2$ component 
observed~\cite{baselmans}.
However the amplitude of higher order components in the magnetic
Josephson junctions was long considered too low for being
observed.
Note now that at $T^*$ the coefficient of the first harmonics
($I_1$) vanishes and, therefore, the higher harmonics become
dominant.
In Ref.~\cite{sellier2} the measured critical current (the maximum
of the absolute value of the current-phase relation) does not
vanish at $T^*$, but passes through a minimum.
This fact, together with the observation of fractional Shapiro
steps, indicates that the observed current is in fact the $I_2$
component.
A phenomenological decoherence time has been proposed in Ref.~\cite{melin}
to fit both $I_1$ and $I_2$ amplitudes at zero temperature.
On the other hand no $I_2$ component was detected in Ref.~\cite{frolov}.

In this Letter we develop a microscopic theory enabling the quantitative
derivation of the full current-phase relation in the regimes
corresponding to actual experiments of Refs.~\cite{frolov,
sellier,sellier2}.
%
%For this purpose we generalize the approach of
%Ref.[\onlinecite{ryazanovbuzdin}] which assumes uniaxial magnetic
%disorder, and 
%
We solve the resulting equations numerically without restriction on the 
values of the parameters.
Extracting the exchange field and magnetic scattering time from
the published data on the temperature dependence of $I_c$, we
estimate the expected magnitude of $I_2$.  
We find that the predicted values agree favorably with those observed in
Ref.~[\onlinecite{sellier2}], while the expected magnitude of
$I_2$ for the sample of Ref.~\cite{frolov} is too small to be
observable.
Note that thickness inhomogeneities in the ferromagnetic layer could give 
an alternative explanation~\cite{frolov2} of the experimental result of Sellier 
{\it et al.}~\cite{sellier2}.
A powerful and microscopic approach to  superconductivity in
disordered metals is offered by the quasiclassical theory in a
form described in \cite{kopnin,bergeret}.
The theory can also describe ferromagnetism, by inclusion of
an exchange field acting on conduction electrons.
This was done for instance in Refs.~\cite{alexander,demler}, where
the spin-orbit coupling with impurities was also included.
However, for the weak ferromagnet Cu$_x$Ni$_{1-x}$ used in
experiments of~\cite{frolov,ryazanov,sellier,sellier2}, the
spin-orbit coupling is expected to play a minor role.
The more important effect should come from the strong
inhomogeneities of the magnetic field on both the microscopic-
(magnetic impurities) and mesoscopic scales (randomly oriented
magnetic domains).  Our theory takes this effect into account.

The metallic ferromagnet is described by the following Hamiltonian:
\begin{equation} 
%\label{eq:hamiltonian}
\nonumber
H=\int \! d\bm{r}
\sum_{ss'}
\psi_s^\dagger
\left[
    \left(
        -\frac{\bm{\nabla}^2}{2m}
        -\mu
        +U
    \right)\delta_{ss'}
    -\bm{h}.\bm{\sigma}_{ss'}
\right] \psi_{s'},
\end{equation}
where $\psi_s(\bm{r})$ and $\psi^\dagger_s(\bm{r}) $ are
annihilation and creation operators for electrons having spin
projection $s$ along the $\hat{z}$ direction, $m$ is the effective
electron mass, and $\mu$ is the Fermi energy ($\hbar=1$).
The disorder potential $U(\bm{r})$ describes the interaction of
electrons with nonmagnetic impurities and is characterized by the
correlation function: 
$\overline{U(\bm{r})U(\bm{r}')}=\delta(\bm{r}-\bm{r}')/(2\pi \nu \tau)$,
%
%\begin{equation}
%  \label{Uave}
%  \overline{U(\bm{r})U(\bm{r}')}=\frac{1}{2\pi \nu \tau}\delta(\bm{r}-\bm{r}'),
%\end{equation}
%
where $\tau$ is the elastic mean free time and $\nu$ is the density of
state at the Fermi level per spin.
The upper bar stands for disorder averaging.
The exchange field $\bm{h}(\bm{r})$ acting on the electron spins may
originate, for instance, from contact interaction between conduction
electrons and localized impurity spins.
We do not consider the question of the microscopic origin of
$\bm{h}(\bm{r})$, but restrict ourselves to setting its
statistical properties only.
Namely, we take its average to be spatially uniform:
$\overline{\bm{h}(\bm{r})} = h \, \hat{z}$, with $h$ proportional
to the magnetization of the ferromagnet.
The fluctuating part is characterized by correlation functions:
\begin{equation}
  \label{have}
\overline{
    (h_\alpha(\bm{r})-\overline{h}_\alpha)
    (h_\beta(\bm{r}')-\overline{h}_\beta)
        }
=
\frac{1}{2\pi \nu \tau_m^\alpha}
\delta_{\alpha\beta}
\delta(\bm{r}-\bm{r}'),
\end{equation}
for $\alpha,\beta=x,y,z$.
Here $\tau_m^\alpha$ characterizes mean free time due to magnetic
impurities.
In the following, we also assume rotational symmetry around $\hat z$,
thus $\tau_m^x=\tau_m^y$.

In order to describe the proximity effect in the ferromagnet, it
is convenient to introduce thermal Green's functions
%
%\begin{equation}
%\label{Gdef}
%\mathcal{G}_{ns,n's'}(\bm{r},\bm{r}',\tau)
%=
%-\langle
%    T_\tau
%    \Psi_{ns}(\bm{r},\tau)
%    \Psi_{n's'}^\dagger(\bm{r}',0)
%\rangle
%\end{equation}
%
$\mathcal{G}_{ns,n's'}(\bm{r},\bm{r}',\tau)
=
-\langle
    T_\tau
    \Psi_{ns}(\bm{r},\tau)
    \Psi_{n's'}^\dagger(\bm{r}',0)
\rangle$
in the Nambu($n$)-spin($s$) space, where $\Psi_{1s}=\psi_s$ and
$\Psi_{2s}=\psi^\dagger_{-s}$.
The equation of motion for the Matsubara-transformed
disorder-averaged Green's function, $\overline{\mathcal{G}}$, is
derived from the Hamiltonian 
%(\ref{eq:hamiltonian}) 
and reads:
\begin{equation}
\left[
    i\omega_n
    - \left(
            -\frac{\bm{\nabla}^2}{2m}-\mu-h\sigma_z
            \right) \tau_z
    -\Sigma_1-\Sigma_2
\right]
\overline{\mathcal{G}}
=
\hat{1}.
\end{equation}
Here, $\omega_n$ are Matsubara frequencies at temperature $T$.
%
%The self-energies $\Sigma_1$
%and $\Sigma_2$ are due to nonmagnetic and magnetic disorder,
%respectively.
%
%Using Eqs. (\ref{Uave}) and (\ref{have}) we derive the self-consistent
%equations for $\Sigma_1$ and $\Sigma_2$: 
%
%\begin{eqnarray*}
%    \Sigma_1(\bm{r},\omega_n)
%&=&
%    (2\pi \nu \tau)^{-1}
%    \tau_z
%    \overline{\mathcal{G}}(\bm{r},\bm{r},\omega_n)
%    \tau_z
%\\
%    \Sigma_2(\bm{r},\omega_n)
%&=&
%    \sum_{\alpha=x,y,z}
%    (2\pi \nu \tau_m^\alpha)^{-1}
%    \tau_z S_\alpha
%    \overline{\mathcal{G}}(\bm{r},\bm{r},\omega_n)
%    S_\alpha \tau_z \,.
%\end{eqnarray*}
%
The non-magnetic disorder leads to the usual forme of the self energy 
$\Sigma_1(\bm{r},\omega_n) =
    (2\pi \nu \tau)^{-1}
    \tau_z
    \overline{\mathcal{G}}(\bm{r},\bm{r},\omega_n)
    \tau_z.$
The magnetic disorder gives instead:
\begin{equation}
\Sigma_2(\bm{r},\omega_n)
    \sum_{\alpha=x,y,z}
    (2\pi \nu \tau_m^\alpha)^{-1}
    \tau_z S_\alpha
    \overline{\mathcal{G}}(\bm{r},\bm{r},\omega_n)
    S_\alpha \tau_z.
\end{equation}
Here $\bm{S}=(\sigma_x,\sigma_y,\sigma_z\tau_z)$, $\sigma_\alpha$ and
$\tau_\alpha$ are Pauli matrices in spin and Nambu spaces,
respectively.

Now we define the quasiclassical Green's function
$g(\bm{r},\omega_n ) = (i/\pi\nu) \tau_z
\overline{\mathcal{G}}(\bm{r},\bm{r},\omega_n)$
%
%\begin{equation}
%g(\bm{r},\omega_n ) = \frac{i}{\nu\pi} \tau_z
%\overline{\mathcal{G}}(\bm{r},\bm{r},\omega_n) \, ,
%\end{equation}
%
which, in the diffusive limit, obeys the equation
\begin{equation} \label{eq:usadel}
-D\bm{\nabla} (g \bm{\nabla} g)
+
[
    \omega_n \tau_z
    -ih\tau_z\sigma_z
    + \sum_\alpha
    \frac{1}{2\tau_m^\alpha} S_\alpha g S_\alpha,
    g
]=0,
\end{equation}
with $D$ the diffusion coefficient, and the normalization condition $g^2=1$.
Symmetry properties of the Hamiltonian further constrain the form of $g$.
Specifically:
({\em i}-a) By the invariance under the rotation around the
average magnetization axis $\hat{z}$, we find that $g$ is block
diagonal in spin space;
we thus define the two matrices in Nambu space, $g_+$ and $g_-$, as the
two non-vanishing upper and lower components, respectively.
({\em i}-b) From the invariance under the rotation over an angle
$\pi$ around the $\hat x$ (or $ \hat y$) axis and simultaneously
change of sign of $h$, we find $ g_+(h) = \tau_zg_-(-h) \tau_z$.
({\em ii}) By time-reversal symmetry, we find
$g_\pm(\bm{r},\omega_n)=-\tau_z g_\pm(\bm{r},-\omega_n)^\dagger \tau_z$.
Finally the 16 correlation functions introduced in Eq.~(\ref{Gdef})
are not independent, since the representation is redundant.
This gives: ({\em iv})
$g_+(\bm{r},\omega_n)=-\tau_x g_-(\bm{r},\omega_n)^* \tau_x$.

Exploiting these properties, we parameterize the complete Green's
function in terms of $g_+$, which in its turn is completely
determined by complex functions, $\theta$ and $\eta$:
\begin{equation}
g_+=
\left(
\begin{array}{cc}
\cos\theta & \sin\theta e^{i\eta} \\
\sin\theta e^{-i\eta} & -\cos\theta
\end{array}
\right) \, .
\end{equation}
Then, Eq. (\ref{eq:usadel}) yields:
%
%\begin{widetext}
\begin{subequations} \label{eq:diffeq}
\begin{eqnarray}
0&=&
    D   \nabla (\sin^2\theta\nabla\eta)
    +\frac{2}{\tau_m^x}
    \sin\theta
    \sin\theta^*
    \sin(\eta-\eta^*)
\\
0&=&
    -D\nabla^2\theta
    +D\cos\theta\sin\theta(\nabla\eta)^2 \nonumber \\
&&
    +2(\omega_n-ih)\sin\theta
    +\frac{2}{\tau_m^z} \sin\theta\cos\theta \nonumber \\
&&
    +\frac{2}{\tau_m^x}
    [
    \sin\theta \cos\theta^*
    +\cos\theta \sin\theta^* \cos(\eta-\eta^*)
    ]
\end{eqnarray}
\end{subequations}
%\end{widetext}
%
These equations constitute the main analytical result of our work.
Note that in non ferromagnetic superconductors, symmetry properties
({\em i}-b) and ({\em iv})
for $h=0$ imply that $\theta$ and $\eta$ are real.
Then Eqs. (\ref{eq:diffeq}) only depend on the effective magnetic
scattering time $1/\tau_m=1/\tau_m^z+2/\tau_m^x$,
in agreement with Abrikosov-Gor'kov theory for magnetic impurities
\cite{abrikosov}.
By contrast, in ferromagnetic superconductors, magnetic disorder can be
characterized by two scattering times: $\tau_m^x=\tau_m^y$ and $\tau_m^z$
\cite{gorkov,fulde}.
In Ref.~\cite{ryazanovbuzdin} the uniaxial disorder was considered.
In our notation this corresponds to $\tau_m^x=\tau_m^y=\infty$,
$\tau_m^z=\tau_m$.
This hypothesis simplifies greatly the solution of
Eq.~(\ref{eq:diffeq}), since $\theta$ and $\eta$ are no more coupled
to $\theta^*$ and $\eta^*$.
The physical reason for this simplification is that
magnetic scattering does not couple the spin up and spin
down populations.
However it seems more realistic that the magnetic disorder is
also able to flip the spin of conduction electrons.
In this sense the opposite
limit is to consider a completely isotropic disorder:
$\tau_m^x=\tau_m^y=\tau_m^z = 3 \tau_m$.
In the following we thus concentrate on this case.
We shall also discuss briefly the uniaxial one for comparison.

\begin{table}
\begin{tabular}{|c|c|c|c|c|c|}
\hline
model   & order &  $h$ (meV) & $ 1/h\tau_m$  & $\rho$ (\%)
& $R I_2$ (nV) \\
\hline
\hline
                            & 1 & 4.6  & 9.3 &          25 & 1.08 \\
$\tau_m^\alpha = 3\tau_m$   & 2 & 49  & 1.1 & \phantom{0}8 & 0.25 \\
(isotropic)                            & 3 & 83  & 0.3 &           10 & 0.07 \\
\hline
$\tau_m^z = \tau_m $        & 1 & 15  & 2.7 &           12 & 0.10 \\
$\tau_m^x = \infty $        & 2 & 61  & 0.8 &  \phantom{0}0 & 0.16 \\
(uniaxial)        & 3 & 86  & 0.2 &  \phantom{0}8 & 0.08 \\
\hline
\end{tabular}
\caption{
\label{FitTable} 
Parameters of the fit (shown in Fig.~\ref{FigFit} 
for the isotropic case). 
We took for the sample $\Delta=1.3$ meV,
$D/L^2=1.13$ meV for $L=17$ nm, $T^*$=1.1 K. $\rho=\sigma_f/(2\sigma_f+\gamma_b
L)$ is the ratio of the barrier resistance to the total resistance $R$
of the junction in its normal state.
}
\end{table}

In order to determine the Josephson relation, we assume that the
ferromagnet is a layer of the length $L$ along the $\hat{x}$-axis.
We thus need to solve Eqs. (\ref{eq:diffeq}) with appropriate boundary
conditions \cite{kupryanov} at $x=\pm L/2$:
\begin{subequations}
\begin{eqnarray}
    \sin \theta\nabla\eta
&=&
    \mp
    \frac{\gamma_b}{\sigma_f}
    \frac{\Delta}{\sqrt{\omega_n^2+\Delta^2}}\sin(\eta\mp\frac{\phi}{2})
\,,
\\
    \nabla \theta
&=&
    \mp
    \frac{\gamma_b}{\sigma_f}
    \frac{\omega_n\sin\theta-
      \Delta \cos\theta\cos(\eta\mp\frac{\phi}{2})}
     {\sqrt{\omega_n^2+\Delta^2}} \,.
\end{eqnarray}
\end{subequations}
Here, $\sigma_f$ is the conductivity of the ferromagnet, and
$\gamma_b$ is the barrier resistance per unit area at the contacts
 (taken to be identical for simplicity)
between the ferromagnet and  the superconducting leads.
We also assume that temperature dependence of the
superconducting gap in the leads, $\Delta$, is simply given by conventional BCS theory.
The supercurrent is given by
\begin{equation}
\label{currentdef}
  I
  =
  \frac{ 2\pi G_f L T}{e}\sum_{\omega_n>0}
  \mathrm{Re} \left[\sin^2\theta\nabla\eta\right],
\end{equation}
where $G_f$ is the conductance of the ferromagnetic metal.

We use now  the above equations for fitting the experimental data
of Ref. \cite{sellier2,frolov}.
For a given set of parameters, namely the Thouless energy $D/L^2$,
$\Delta(T=0)$, $T$, $\gamma_b$, $h$, $\tau_m$, we
obtain the current-phase relation by solving numerically the
system of differential equations (\ref{eq:diffeq}) and
calculating the current through Eq. (\ref{currentdef}).
One can then extract the first two harmonics, $I_1$ and $I_2$.
(Higher harmonics near $T^*$ are much smaller than $I_2$).
We begin with the data of Ref.~\cite{sellier2}, Fig. 2,
concerning a sample with a ferromagnetic layer of $L=17$ nm.
The length $L$, the superconducting gap, and the temperature
are known experimentally.
The interface resistance is more difficult to measure.
The authors of Ref. \cite{sellier2} give an estimate of 30\% of the
total resistance of the junction in its normal state, $R$ \cite{sellier}.

\begin{figure}
\centerline{\psfig{file=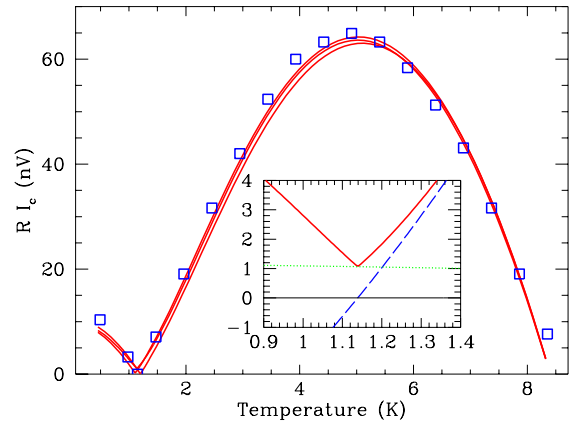,angle=-90,width= 7.8cm}}
\caption{
\label{FigFit} Fit to experimental data for the critical
current of Ref. \cite{sellier2} (boxes), for the three solutions
reported in the table of the isotropic model.
The three solutions gives nearly indistinguishable curves.
Inset: temperature dependence near $T^*$ of the calculated $I_c={\rm
max}_\phi(|I(\phi)|)$, $I_1$ (dashed), and $I_2$ (dotted), for the
first solution.  The minimum of $I_c$ coincides with $I_2$.  }
\end{figure}

For a given value of $\gamma_b$ one can find the pairs of values ($h$,
$\tau$) that satisfy the two equations $I_1(h,\tau_m,T^*)=0$ and
$I_1(h,\tau_m,T_1)=I_1^{exp}$, where $T_1$ is a temperature different
from $T^*$ and $I_1^{exp}$ is the corresponding experimental value for
$I_1$.
We find that only three pairs of values satisfy this constraint.
We order them by increasing value of $h$.
One can show that the $n$-th solution refers to a junction where, as a
function of the length of the sample, other $n-1$ zeros are predicted
for $L<17$ nm.
We optimize then this first estimate of the parameters by including
$\gamma_b$ as a fitting parameter for the full experimental curve.
The solutions are given in Table \ref{FitTable}.
For comparison fitting parameters for uniaxial magnetic disorder
are also given.

In all cases we have a good fit to data (see Fig. \ref{FigFit}).
Thus, the quality of the fit is not a sufficient criterium to discriminate
between the three possibilities (for each model).
One argument in favor of the first solution (for isotropic model of
magnetic disorder) is the agreement of the fitting parameter
$\gamma_b$ with the estimated value in the experiment.
A second one is the dependence of $T^*$ on the
length that increases with the order of the solution.
The range of $L$ for which $0<T^*<T_c$ is
about 1 nm for the first solution,
and about $0.4-0.3$ nm for the second and third one.
The first case compares better with the experiment
\cite{sellier2}, where incertitude on $L$ is about $1$ nm,
and $\pi$-contacts were observed for $L=17-19$ nm \cite{sellier}.

\begin{figure}
\centerline{\psfig{file=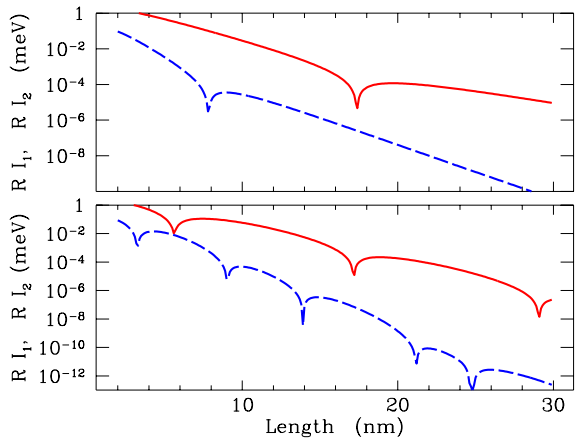,angle=-90,width= 8cm}}
\caption{\label{LengthDep} Length dependence at $T=4.2$ K of $I_c$ 
(full line) and $I_2$ (dashed line) for the
parameters obtained with the fit with the isotropic model.  We show
the solution 1 (upper panel) and 2 (lower panel). The zero at $L=17$ nm is
visible in both cases. Solution 2 displays
a second zero for $L<17$ nm.
}
\end{figure}

We consider now the second component, $I_2$.

In Ref. \cite{sellier2} the minimum value of $R I_2$ is $0.5$ nV,
and it falls between the first and the second predicted value
for isotropic model (cf. Tab. \ref{FitTable}).
In both cases we thus find that the amplitude of the second harmonics
is compatible with the observed one.
We also find a strong temperature dependence of $I_2$,
the values presented in the table at $T^*=1.1$ K,
are reduced by a factor 10 at 5 K.
This dependence can explain the much smaller value for $R I_2$ observed
in the 19 nm sample of Ref. \cite{sellier2}.
In comparison, the uniaxial model gives a much smaller value for
$R I_2 \approx 0.1$ nV for all three solutions.

We repeated the fitting procedure on the data of Frolov {\em et al.}
\cite{frolov}, where no second harmonics is observed
at $T^*$.
We found again that the data can be compatible with either a
second or a first zero, but in both cases $R I_2 < 10^{-10}$ mV,
thus below the observation threshold.

We finally discuss the length dependence of the first and second
harmonics for the fitted values of the parameters (see
Fig. \ref{LengthDep}).
As anticipated, $I_1(L)$ displays an oscillating behavior.
One can clearly see in Fig. \ref{LengthDep} that $I_1$ for
solution 1 and 2 vanishes once and twice, respectively, for
$L\leq 17$ nm (solution 3 is not shown).
A more unexpected result is the oscillatory behavior for $I_2(L)$,
that shows a remarkable doubled periodicity with respect to $I_1(L)$:
Between two zeros of the first harmonics we {\em always}
observed two zeros of the second harmonics.
This means that the sign of $I_2$ remains always
positive when $I_1$ vanishes.
Therefore we find that the transition from $0$- to $\pi$-contact is
always discontinuous.
We cannot rule out, however, that $I_2$ may be negative at $T^*$ in
some other region of the parameters space.
This would imply that the transition from $0$ to $\pi$-contact
is continuous as a function of the temperature \cite{buzdin2}.

In conclusion, we have presented a development of the
quasiclassical theory of superconductivity taking into account
magnetic scattering in the presence of an exchange field.
We have used our model to extract the exchange field and
scattering times from the temperature dependence of the critical
current in superconductor/ferromagnet/superconductor junctions.
With these parameters we have calculated the second harmonics at
the vanishing value of the first component which agree favorably
with the experimental findings.

{\em Note added}: after the completion of this work we became aware of
related work by A. Buzdin \cite{buzdinU} where $I_2$ is calculated 
near the critical temperature.

We are pleased to thank A. Buzdin for illuminating and important
discussions.
This work was supported in part by the U.S. Department of Energy
office of Science under the contract W-31-109-Eng-38.

\end{document}